\documentclass[twocolumn,prd,superscriptaddress,amssymb,nofootinbib,amsmath,nobibnotes,aps,prlgroupedaddress]{revtex4-2} 

\usepackage{graphicx}  
\usepackage{dcolumn}   
\usepackage{bm}        
\usepackage{amssymb}   
\usepackage[]{amsmath,amssymb}
\usepackage{graphics,epsfig}
\usepackage{float}
\usepackage{booktabs}
\usepackage{mathtools}
\usepackage{esint}
\usepackage[utf8]{inputenc}
\usepackage[english]{babel}
\usepackage{mathtools}
\usepackage{xcolor}
\usepackage{setspace} 
\usepackage{indentfirst}  
\usepackage{cancel}  
\usepackage{hyperref}  
\hypersetup{colorlinks=true, linkcolor=blue } 
\usepackage{tcolorbox}   
\usepackage{ulem} 

\usepackage{orcidlink}

 \def\b{\beta} \def\g{\gamma} \def\d{\delta} 
\def\m{\mu} \def\n{\nu} \def\r{\rho} \def\s{\sigma}  \def\o{\omega}
\def\O{\Omega}\def\L{\Lambda}\def\S{\Sigma}\def\T{\Theta}\def\t{\tau}
\def\N{\nabla} \def\h{\hat} \def\OA{\O_{\rm A}} \def\Ocdm{\O_{\rm cdm}} \def\mA{m_{\rm A}}

\date{\today}
\begin{document}
\title{Spin-1 Ultralight Dark Matter under Cosmological Scrutiny: Mass Constraints from CMB and Distance Probes}

\author{Guadalupe A. Acu\~na \orcidlink{0009-0009-2608-7718}} 
\email{gi.ahumadaacuna@df.uba.ar}
\affiliation{Universidad de Buenos Aires, Facultad de Ciencias Exactas y Naturales, Departamento de Física. Buenos Aires, Argentina.} 
\affiliation{CONICET - Universidad de Buenos Aires, Instituto de Física de Buenos Aires (IFIBA). Buenos Aires, Argentina}
\affiliation{Instituto de F\'{i}sica, Universidade Federal do Rio Grande do Sul, 91501-970 Porto Alegre RS, Brazil} 

\author{Tomas F. Chase \orcidlink{0009-0001-0286-2136}} 
\email{tferreirachase@df.uba.ar}
\affiliation{Universidad de Buenos Aires, Facultad de Ciencias Exactas y Naturales, Departamento de Física. Buenos Aires, Argentina.} 
\affiliation{CONICET - Universidad de Buenos Aires, Instituto de Física de Buenos Aires (IFIBA). Buenos Aires, Argentina}

\author{Diana L\'opez  Nacir \orcidlink{0000-0003-4398-1147}} 
\email{dnacir@df.uba.ar}
\affiliation{Universidad de Buenos Aires, Facultad de Ciencias Exactas y Naturales, Departamento de Física. Buenos Aires, Argentina.} 
\affiliation{CONICET - Universidad de Buenos Aires, Instituto de Física de Buenos Aires (IFIBA). Buenos Aires, Argentina}

\author{Rafael C. Nunes \orcidlink{0000-0002-8432-5616}}  
\email{costa.nunes@ufrgs.br}
\affiliation{Instituto de F\'{i}sica, Universidade Federal do Rio Grande do Sul, 91501-970 Porto Alegre RS, Brazil} 
\affiliation{Divisão de Astrofísica, Instituto Nacional de Pesquisas Espaciais, Avenida dos Astronautas 1758, São José dos Campos, 12227-010, São Paulo, Brazil.}


\begin{abstract}
We present cosmological constraints on spin-1 ultralight dark matter, described by a vector field (VFDM) with mass $m_{\rm A}$, using Planck CMB data and geometrical probes from BAO and SNIa. A key theoretical result is the derivation of the full CMB temperature covariance matrix, including both diagonal and off-diagonal anisotropic contributions induced by the preferred direction of the background vector field. We first constrain the model using the diagonal part of the covariance, together with CMB lensing; the off-diagonal terms, which couple multipoles with $\Delta\ell\in\{2,4\}$, could bias lensing reconstruction, but only at very low multipoles ($L=\{2,4\}$) not included in the Planck likelihood. We consider both a pure VFDM scenario and a mixed VFDM+CDM scenario, characterized by the fraction $f=\Omega_{\rm A}/(\Omega_{\rm A}+\Omega_{\rm cdm})$, obtaining $\log_{10}(m_{\rm A}/\mathrm{eV})>-24.07$ (95\% C.L.) in the pure case, and a clear correlation between $f$ and $m_{\rm A}$ in the mixed case, with smaller fractions allowing lighter masses; standard cosmological parameters remain fully consistent with $\Lambda$CDM. For the off-diagonal contributions, we derive the corresponding Bipolar Spherical Harmonic (BipoSH) coefficients and predict their amplitude using our best-fit and bounds. While the anisotropic signal is difficult to detect in the pure VFDM scenario with current Planck data, mixed VFDM+CDM models can produce signals at, or above, Planck sensitivity over a range of multipoles, motivating dedicated searches for this characteristic signature.
\end{abstract}
\maketitle


\section{Introduction}\label{sec:introduction}

The $\Lambda$CDM cosmological model has proven to be remarkably successful in 
describing the large-scale structure of the Universe, the anisotropies of 
the cosmic microwave background (CMB), and the late-time expansion history 
\cite{Dodelson:2020bqr,Planck:2018vyg, DESI:2025zgx, Weinberg:2013agg, SDSS:2003eyi, SupernovaSearchTeam:1998fmf}. A key component of this framework 
is cold dark matter (CDM), whose gravitational effects are well established 
due to a large amount of observations, including galaxy rotation curves, 
gravitational lensing, CMB anisotropies, and large-scale structure 
\cite{Bertone:2016nfn, Bertone:2004pz}. Despite its observational successes, 
the microscopic nature of dark matter remains one of the most pressing open 
questions in modern physics.

While $\Lambda$CDM provides an excellent description of the Universe on 
large scales, it faces persistent challenges on sub-galactic scales \cite{DelPopolo:2016emo}. 
Cosmological N-body simulations predict dark matter halos with cuspy central 
density profiles, in tension with the cored profiles inferred from galaxy rotation curves (the \textit{cusp-core problem})  \cite{Oh_2011, Navarro:1995iw}. 
Additionally, simulations predict a far larger number of low-mass satellite 
galaxies than are observed around the Milky Way (\textit{missing satellites 
problem}), and the most massive predicted subhalos appear too dense to host 
the observed dwarf spheroidal galaxies (\textit{too-big-to-fail problem}) 
\cite{Boylan-Kolchin:2011qkt, Boylan-Kolchin:2011lmk, Klypin:1999uc}. Although baryonic feedback 
mechanisms can alleviate some of these tensions \cite{Weinberg:2013aya}, 
they motivate the exploration of dark matter candidates with novel 
small-scale properties.

On the other hand, beyond the traditional small-scale structure debates, statistically significant tensions between different cosmological datasets have emerged in recent years, potentially signaling the need for new physics beyond the standard model. The most prominent of these is the Hubble tension: a discrepancy exceeding 5$\sigma$--7$\sigma$ \cite{H0DN:2025lyy} between the value of the Hubble constant, $H_0$, inferred from early-universe measurements and that obtained from local distance ladder observations (see \cite{CosmoVerseNetwork:2025alb,Perivolaropoulos:2021jda} for a review). A second major discrepancy involves the $S_8$ parameter \cite{Nunes:2021ipq,Pantos:2026koc}, along with several other tensions of more modest statistical significance that have also appeared in recent years \cite{CosmoVerseNetwork:2025alb}. Taken together, these tensions—arising from multiple independent and complementary datasets—have motivated investigations into extensions of the standard $\Lambda$CDM model \cite{CosmoVerseNetwork:2025alb}.

Ultralight dark matter (ULDM), consisting of bosonic particles with masses 
$m \lesssim$ eV, has attracted significant attention as an 
alternative or complement to standard CDM \cite{Ferreira:2020fam}. The key feature of ULDM is its macroscopic de Broglie wavelength, which suppresses structure formation below a characteristic Jeans scale and produces cored halo profiles supported 
by an effective quantum pressure \cite{Hu:2000ke, Guzman:2004wj, Hui:2016ltb}. This framework is theoretically well motivated: ultralight bosonic fields arise
generically in string compactifications and other beyond-the-Standard-Model
scenarios. While ultralight scalars have been extensively studied in the
context of axion dark matter
\cite{Arvanitaki:2009fg,Marsh:2015xka,Weinberg:1977ma,Wilczek:1977pj,Preskill:1982cy},
ultralight vector fields can also naturally emerge from hidden-sector
$U(1)$ gauge symmetries and provide viable dark matter candidates
\cite{Goodsell:2009xc,Nelson:2011sf,Arias:2012az,Goodsell:2009xc}. 

In the case where the presence of a homogeneous ultralight  field accounts for the entire background dark matter abundance, cosmological observations place stringent constraints on ULDM models. Such background scalar fields can be produced by the so-called misalignment mechanisms (see for instance \cite{Marsh:2015xka}).
For this kind of scalar ULDM models, measurements of the Lyman-$\alpha$ forest require masses above $m \gtrsim 10^{-21}\,\mathrm{eV}$ \cite{Irsic:2017yje,Armengaud:2017nkf}, with more recent analyses obtaining even stronger bounds of $m \gtrsim 2\times 10^{-20}\,\mathrm{eV}$   \cite{Rogers:2020ltq}. Likewise, Planck CMB data combined with large-scale structure observations exclude masses below $m \sim 10^{-24}\,\mathrm{eV}$. For scenarios in which the ultralight component coexists with standard CDM, the same datasets restrict the allowed ULDM fraction to sub-percent levels for masses in the range $10^{-32} \,\mathrm{eV}\lesssim m \lesssim 10^{-25.5} \,\mathrm{eV}$ \cite{Hlozek:2014lca}. These results indicate that mixed dark matter scenarios, in which an ultralight component contributes only a fraction of the total dark matter abundance, remain phenomenologically viable. In this work, we investigate the analogous possibility for vector-field dark matter, allowing it to constitute a fraction of the total dark matter density.

Beyond the scalar case, ULDM can also be realized as a spin-1 (vector) 
field, described by the Proca action. Homogeneous background vector fields can be produced during inflation, thereby accounting for their relic abundance today \cite{Nakayama:2019rhg,Kitajima:2023fun}. The cosmological phenomenology of 
vector field dark matter (VFDM) differs significantly from that of its scalar 
counterpart. Most notably, a homogeneous background vector field breaks the 
spatial isotropy of the Universe, requiring a Bianchi~I metric rather than 
the standard FLRW background \cite{Chase:2023puj}. As a consequence, the 
energy density of VFDM scales as $\rho_{\rm A} \propto a^{-4}$ before the 
onset of oscillations—in contrast to the approximately constant behavior of 
scalar ULDM in this regime—and the model sources a background anisotropic 
shear that evolves according to the Einstein equations 
\cite{Chase:2023puj, Chase:2024wsq}. These distinctive features generate 
off-diagonal correlations in the CMB covariance matrix, breaking 
statistical isotropy in a characteristic pattern dictated by the symmetry of 
the vector field, as will be shown in the present work.

The theoretical framework for VFDM has been developed and implemented in a 
modified version of the \texttt{CLASS} Boltzmann solver in 
\cite{Chase:2023puj, Chase:2024wsq,Chase:2026qxs}\footnote{See \url{https://github.com/classULDM/class.VFDM}.}. However, a systematic Bayesian 
analysis of the VFDM parameter space using state-of-the-art cosmological 
datasets has not yet been performed. In this work, we fill this gap by 
carrying out Markov Chain Monte Carlo (MCMC) analyses using \texttt{MontePython} 
\cite{MP_ref1, MP_ref2} with three complementary dataset combinations: 
Planck 2018 CMB temperature and polarization data \cite{Planck:2018vyg, 
Planck:2019nip}, baryon acoustic oscillation measurements from the DESI 
second data release \cite{DESI:2025zgx, DESI:2025zpo}, and Type Ia 
supernovae from the PantheonPlus compilation \cite{Brout:2022vxf}. We 
consider two scenarios: a \textit{pure} VFDM model in which the vector field 
constitutes all of the dark matter, and a \textit{mixed} scenario in which 
VFDM coexists with standard CDM. In both cases, we place constraints on the 
vector field mass $m_{\rm A}$ and its contribution to the dark matter 
abundance, and compare the resulting fits with those obtained in the $\Lambda$CDM model.

This paper is organized as follows. In Section~\ref{sec:model} we describe 
the VFDM model and its cosmological signatures, including the background 
equations and the full covariance matrix of temperature anisotropies. 
Section~\ref{sec:methodology} presents the datasets and the 
methodology employed. The main results are reported and discussed in 
Section~\ref{sec:results}. In Section~\ref{sec:offdiag} we discuss on the relevance of the off-diagonal elements of the CMB covariance matrix to constrain the model. We present our conclusions in 
Section~\ref{sec:conclusions}. In App. \ref{app:delta_Cl} we asses the detectability of the anisotropic contributions of the vector field to the diagonal part of the CMB covariance trough a Fisher analysis, and in App. \ref{app:lensing_vfdm} we calculate the multipoles at which our model predicts a spurious lensing-like contribution. 


\section{Spin-1 ULDM Model}\label{sec:model}

The Spin-1 ULDM particle considered here is described by a vector field $A^{\m}(\t,\vec{x})$ within the framework of General Relativity, and whose action is given by 
\begin{equation}\label{eq:action_vf}
    S = -\int d\t\,dx^3\,\sqrt{-g}\left[\frac{1}{4} F^{\m\n}F_{\m\n} + \frac{\mA^2}{2} A^{\m}A_{\m}\right]\, ,
\end{equation}
where $g$ denotes the determinant of the metric, $F_{\m \n} = \N_\m A_\n - \N_\n A_\m$ is the field strength tensor and $\mA$ is the vector field mass. Varying the action with respect to $A_{\m}$ results in the equation of motion for the vector field, given by the Proca equation 
\begin{equation}\label{eq:proca}
    \N_\m F^{\m \n} + \mA^2 A^\n = 0.
\end{equation}

We can study the background evolution of the theory by considering the vector field as a combination of an homogeneous (but not isotropic) background field plus a perturbation, i.e. $A^\m (\t,\vec{x}) = A^\m (\t) +\d A^\m(\t, \vec{x})$. Upon substitution, the equation of motion (Eq. \eqref{eq:proca}) splits into a dynamical equation for the spatial components $\vec{A}$ and a constraint equation for the time component $A_0$. The background vector field solution can be parameterized as $\vec{A} = A(\t) \h A$ (assuming that it points in a given direction), with  $A_0=0$. 

Since the  energy-momentum tensor of the field has a background anisotropic stress, the consistency of the model requires the background spacetime to be an anisotropic Bianchi-I metric \cite{Chase:2023puj}, 
\begin{equation}
    ds^2 = a(\t)\left[-d\t^2 + \gamma_{ij} dx^i dx^j\right]\,,
\end{equation}
which is a generalization of the FLRW (Friedmann-Lema\^itre-Robertson-Walker) spacetime. Here  $a(\t)$ is the scale factor, $\t$ is the conformal time, related to the cosmic time $t$ by $dt = a(\t) d\t$, and $\g_{ij}$ (with $i,j\in \{1,2,3\}$) are the spatial components of the metric, given by $\g_{ij}=e^{-2 \b_i(\t)} \d_{ij}$, with $\b_i$ constrained by $\sum_{i}^{3} \b_i = 0$. The anisotropies are described by the the shear tensor, defined as $\s_{ij} = \tfrac{1}{2} \g'_{ij}$ \cite{Pereira:2007yy}, where prime derivatives are with respect to $\t$.

Due to the anisotropic nature of the background metric, the time derivative of the vector direction $\h A$ does not vanish but is of order $\mathcal{O}(\s_A \equiv \h A_i\h A_j\s^{ij})$. To leading order in $\s_A$ we neglect the time derivatives of $\h A$ and focus on an equation of motion for the vector modulus $A(\t)$. Eq.~\eqref{eq:proca} then reduces to
\begin{equation} \label{eq:eom}
  A''+ \mA^2  a^2A = 0.
\end{equation}
An approximate solution of the second equation can be found in two regimes depending on the relation of the mass and the Hubble rate. When $\mathcal{H} \gg \mA a$ (which is equivalent to $ H \gg \mA $ in cosmic time) the equation can be approximately solved by a growing mode proportional to the scale factor, while for $\mathcal{H} \ll \mA a$ the equation can be solved under a WKB (Wentzel, Kramers, Brillouin) approximation, 
\begin{equation}
    A(\t)\propto\left\{\begin{matrix}
    a & a<a_{\rm {osc}}  \\
    a^{-\tfrac{1}{2}}\cos(\int \mA \ a \, d\t) & a>a_{\rm {osc}}  \\
    \end{matrix}\right. \quad,
\end{equation}
where $a_{\rm {osc}}$ is defined by $\mathcal{H}_{\rm{osc}} = \mA a_{\rm {osc}}$, and corresponds to the time when the field starts oscillating. 

From these solutions we can compute the effective fluid variables, namely the energy density $\r_{\rm A}$, pressure $P_{\rm A}$ and anisotropic stress ${\S^i}_j$, obtained from the energy–momentum tensor. The shear is treated perturbatively, keeping it to first order in the left-hand side of the Einstein equations while neglecting it in the fluid variables.

The background energy density reads
\begin{equation}
    \rho_{\rm A} = \frac{A'{}^{2}+ \mA^2 a^2 A^2}{2 a^2} \propto 
    \begin{cases}
    a^{-4}  & a<a_{\rm {osc}} \\
    a^{-3}  & a>a_{\rm {osc}}
    \end{cases}.
    \label{eq:bckg_energy_density}
\end{equation}

Before oscillations start, the field behaves as a relativistic component with equation of state $w_{\rm A} = 1/3$. Once oscillations set in, the average equation of state approaches that of cold dark matter, $w_{\rm A} = 0$. A notable difference with scalar field dark matter models is that in the vector case the density scales as $a^{-4}$ in the early regime ($a<a_{\rm {osc}}$), whereas for scalar fields the density remains approximately constant before oscillations.

Using the field solution we can also compute the background anisotropic stress of the vector field,
\begin{equation}
    {\S^i}_j =  
        - 6 P_{\rm A} \left(\h{A}_i \h{A}_j - \frac{\g_{ij}}{3} \right),
\end{equation}
whose evolution follows that of the pressure.

   The Einstein equations for   a Bianchi I spacetime  yield to the generalized Friedmann equation,
\begin{equation}
    \mathcal{H}^2 = \frac{\s^2}{6} +  \frac{a^2}{3m_P^2} \rho_{\rm T}\,,
    \label{eq:modified_friedmann}
\end{equation}
where $\rho_T$ is the total energy density and $m_P$ denotes the Planck mass, and the evolution equation for the shear tensor, obtained from the traceless spatial part of Einstein equations,
\begin{equation}
    ({\sigma^i}_j)' + 2 \mathcal{H}\, {\sigma^i}_j = \frac{a^2}{m_P^2} {\Sigma^i}_j .
    \label{eq:metric_shear}
\end{equation}
 The latter can be solved perturbatively assuming negligible anisotropies before some early time $a_{ini}$. Since at leading order we can neglect the time derivatives of the versors, the tensor structure of the metric shear is the same as the one of the source, 
\begin{equation}
    {\s^i}_j =
         \frac{3}{2} \s_A \left(\h A_i \h A_j - \frac{\g_{ij}}{3} \right),
\end{equation}
where $\s_A=\h A_i \h A_j \s^{ij}$, and notice that $\sigma^2=3\sigma_A^2/2$.  Defining the \textit{shear abundance} from Eq. \eqref{eq:modified_friedmann}, we have that $\O_\s = \frac{\s^2}{6\mathcal{H}^ 2}$. At early times, when $a<a_{\rm osc}$, it has been shown in \cite{Chase:2023puj} that this abundance is constant, with a value of 
\begin{equation}\label{eq:shear_const}
    \O_{\s} \simeq 4 \, \O_{\rm{A,0}}^2 \O_{r,0}^{-3/2}\left(\frac{H_0}{\mA}\right) ,
\end{equation}
where $\O_{\rm{A,0}}$ and $\O_{r,0}$ are the  energy budget of VFDM and radiation at the present time, respectively. For later times, the behavior depends on whether the era is dominated by radiation or by matter,
\begin{equation}
    \O_\s \propto  \begin{cases}
    a^{-2}  & a_{\rm {osc}}<a<a_{\rm eq} \\
    a^{-3}  & a>a_{\rm eq}
    \end{cases} \ .
\end{equation}
One of the most restrictive constraints on the shear abundance is imposed by the Big Bang nucleosynthesis (BBN), which gives $\O_{\s} |_{\rm BBN} \leq 10^{-2}$ for a universe without anisotropic sources in the background \cite{Akarsu:2019pwn}, where $\O_{\s} |_{\rm BBN}$ is the shear abundance at BBN ($a_{\rm BBN} \sim 10^{-8}$). Considering that VFDM accounts for a fraction $f$ of the total dark matter in the universe ($\O_{\rm dm}$), that is, $\OA=f \, \O_{\rm dm}$, we can combine the previous constraint with Eq. \eqref{eq:shear_const} to obtain an upper limit on the mass,  
\begin{equation}
    \begin{aligned}
& m_{\rm A} \gtrsim 4 \ f^2\, 10^2  \O_{\rm dm,0}^2 \  \O_{r,0}^{-3 / 2} H_0 \\
& \sim 3 \ f^2 \, 10^{-26} \mathrm{eV}\left(\frac{\O_{\rm dm,0}}{0.26}\right)^2\left(\frac{\O_{r,0}}
{10^{-4}}\right)^{-\frac{3}{2}}\left(\frac{\mathrm{H}_0}{10^{-33} \mathrm{eV}}\right),
\end{aligned}
\end{equation}
where, in the second line, we used approximate values $\O_{\rm dm,0} \sim 0.26, \O_{r,0} \sim 10^{-4}, H_0 \sim 10^{-33} \mathrm{eV}$, estimated from the Planck cosmological parameters. For such parameters we obtain the bound 
 \begin{equation}\label{eq:BBN_constr}
     m_{\rm A} \gtrsim 3f^2 \,  10^{-26}  \mathrm{eV}.
 \end{equation}
As we will see in Section \ref{sec:results}, this bound can be improved by using other dataset combinations.  

An additional lower bound on the vector field mass arises already at the background level. Before the onset of oscillations, the VFDM component behaves as radiation, according to Eq. \eqref{eq:bckg_energy_density}. As shown in Ref.~\cite{Chase:2023puj}, the ratio between the vector and radiation energy densities during radiation domination is approximately constant and given by
\begin{equation}
R_A \equiv \frac{\rho_A}{\rho_r}
\simeq f \Omega_{\rm dm,0} \Omega_{r,0}^{1/4} \left(\frac{H_0}{m_{\rm A}}\right)^{1/2}.
\end{equation}
Therefore, $R_A\propto m_{\rm A}^{-1/2}$, implying that decreasing the vector field mass increases the amount of radiation-like energy present before recombination.
This modifies the early expansion history and, consequently, the sound horizon at decoupling. Although part of this effect can be compensated by changes in late-time cosmological parameters, the additional radiation also affects other CMB observables, such as the epoch of matter-radiation equality and the relative heights of the acoustic peaks. As a result, sufficiently small masses are disfavored by CMB data. This behavior is markedly different from scalar-field dark matter models \cite{Hlozek:2014lca}, where the field energy density remains nearly constant before oscillations and constitutes a negligible fraction of the total energy density in the early Universe.

For a more detailed treatment of the algebraic manipulations presented so far and the implementation of these equations in \texttt{CLASS}, as well as the initial conditions of the fields, we refer the reader to the articles \cite{Chase:2023puj, Chase:2024wsq, Chase:2026qxs}.

\subsection{Full covariance matrix of temperature anisotropies}
The standard decomposition of the temperature perturbation $\Theta(\vec{x},\hat p, \t)$,
\begin{equation} \label{eq:theta_0}
\Theta(\vec{x},\hat p, \t)
  = \sum_{\ell =1}^{\infty} \sum_{m=- \ell} ^{ \ell} 
    a_{\ell m}(\vec{x}, \t)\, Y_{\ell m}(\hat p), 
\end{equation}
 assumes that the only angular dependence enters through the photon momentum direction $\hat p$. The multipoles $a_{\ell m}$ carry the full statistical information of the temperature fluctuations; under the assumption of statistical isotropy, their ensemble average takes the familiar diagonal form
\begin{equation}\label{eq:Cl_standar}
\langle a_{\ell  m} a^{*}_{\ell ' m'} \rangle = C_\ell^{TT} \, \delta_{\ell  \ell'} \delta_{m m'} ,
\end{equation}
where \( C_\ell^{TT} \) is the angular power spectrum of temperature anisotropies, given by 
\begin{equation}\label{eqCl2}
C_\ell^{TT}= 4\pi \int \frac{dk}{k} \, \mathcal{P}_{\mathcal{R}}(k) \, |T_{\ell }(k)|^2 .
\end{equation}
Here $T_{\ell }(k)$ is defined through
 \begin{equation}
 T(k, \mu) = \sum_{\ell} (-i)^\ell (2\ell+1) \mathcal{P}_\ell(\mu) T_\ell (k), 
 \end{equation}
 where $\mu={\hat k}\cdot{\hat p}$, $\mathcal{P}_\ell(\mu)$ denote the Legendre polynomials and $T(k, \mu)\equiv \Theta(k,\mu)/\mathcal{R}(\vec{k})$ is the transfer function at present time. $\mathcal{R}(\vec{k})$ is the primordial curvature perturbation, whose power spectrum $P_\mathcal{R}(k)$ is defined by
\begin{equation}\label{eq:primordials}
 \langle \mathcal{R}(\vec{k})  \mathcal{R}^*(\vec{k}^{\prime})\rangle=\delta(\vec{k}-\vec{k}^{\prime})(2 \pi)^3  P _\mathcal{R}(k),
\end{equation}
and in Eq.~\eqref{eqCl2} we have expressed the result in terms of the dimensionless power spectrum $\mathcal{P}_{\mathcal{R}}(k) = \frac{k^3}{2\pi^2} P_{\mathcal{R}}(k)$.

\begin{figure}
    \centering
    \includegraphics[width=\linewidth]{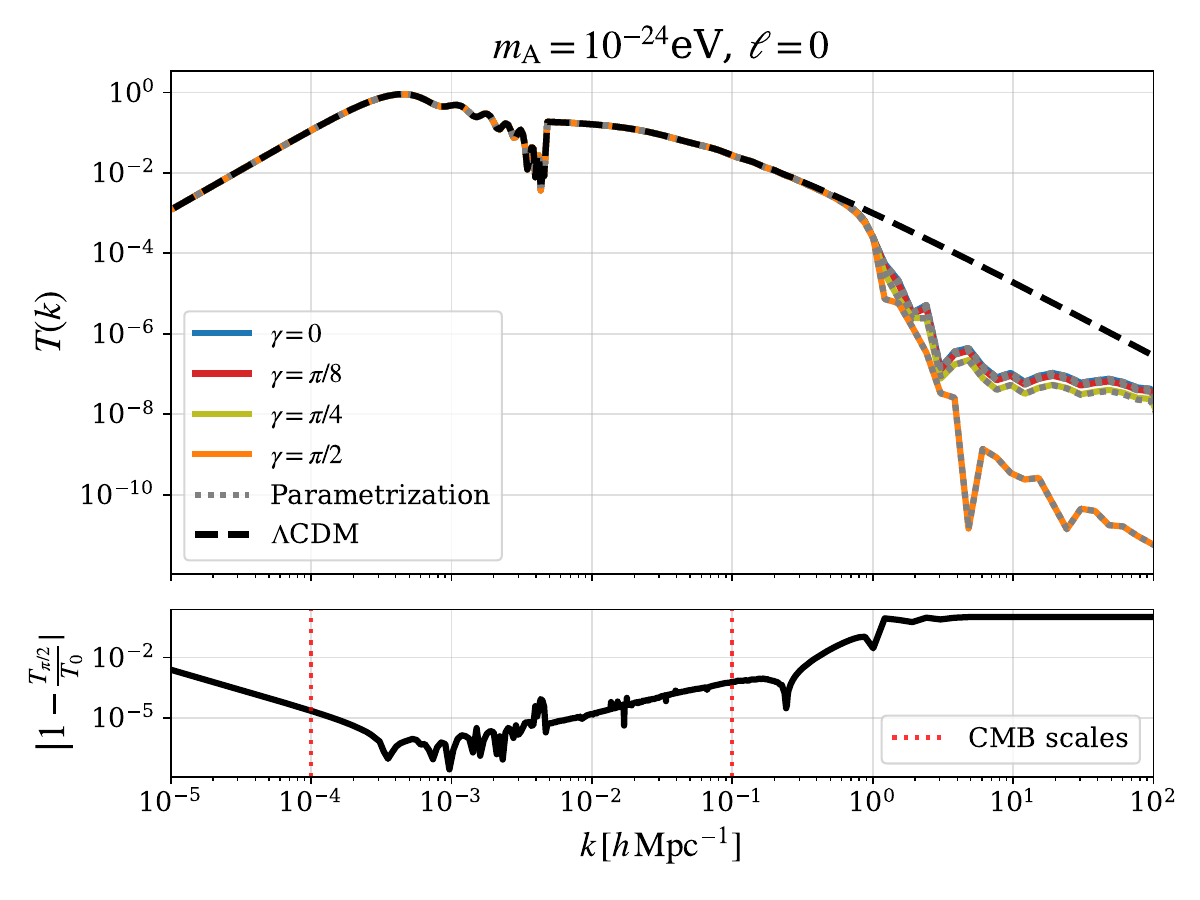}
    \caption{Top panel: photon transfer function for $\ell =0$ and four values of $\g$, computed for $\mA=10^{-24}$ eV with the Planck 2018 best-fit parameters~\cite{Planck:2019nip}. Gray dashed lines show the parametrization of Eq.~\eqref{eq:param_transfer}; black lines show the standard $\Lambda$CDM result.  Bottom panel: relative difference between $T_{\pi/2}$ and $T_{0}$. Red lines delimit the relevant scales probed by the CMB.} 
    \label{fig:transfer_f}
\end{figure}

 However, if the temperature perturbations also depend on the direction of a dark matter vector field ${\h A}$, then $\T = \T (\vec{x},{\h p}, {\h A}, \t)$. If ${\h A}$ is a fixed direction, since the expansion in $Y_{\ell m}$ is only used to describe a continuous angular dependence, the expansion in Eq. \eqref{eq:theta_0} remains valid, but now the coefficients $a_{\ell m}$ contains the anisotropy induced by ${\hat A}$, 
 \begin{equation} \label{eq:theta_A}
\Theta(\vec{x}, {\hat{A}}, {\hat p}, \t)
  = \sum_{\ell ,m} a_{\ell m}(\vec{x}, {\hat A}, \t) \, Y_{\ell m}({\hat p}).
\end{equation}
Let us now find an analogous to Eq. \eqref{eq:Cl_standar} for the VFDM theory. To do so, we first take the Fourier transform of Eq. \eqref{eq:theta_A} (evaluated at $\t=\t_0$, which we will omit in the argument) and compute 
\begin{equation} \label{eq:two-point}
\begin{aligned}
& \left\langle a_{\ell  m}(\vec{x}, {\hat A}) a_{\ell ^{'} m^{\prime}}^*(\vec{x}, {\hat A})\right\rangle= \\
& =\int \frac{d^3 k}{(2\pi)^3} \frac{d^3 k'}{(2\pi)^3} e^{i(\vec{k}-\vec{k}^{\prime}) \cdot \vec{x}} \int d \O d \O^{\prime} Y_{\ell  m}({\hat p}) Y^*_{\ell ^\prime m^\prime}({\hat p}^{\prime}) \, \times\\
&\qquad \qquad \qquad \qquad \qquad \, \left\langle\Theta(\vec{k}, {\hat p}, {\hat A}) \Theta^* (\vec{k}^{\prime}, {\hat p}^{\prime}, {\hat A} )\right\rangle \\
&=\int \frac{d^3 k}{(2\pi)^3} \int d \O d \O^{\prime} {Y}_{\ell m }({\hat p}) {Y}_{\ell ^{\prime} m^{\prime}}^*({\hat p}^{\prime})  T(\vec{k}, {\hat p}, {\hat A}) T^*(\vec{k}, {\hat p}^{\prime}, {\hat A}) P_{\mathcal{R}}(k),
\end{aligned}
\end{equation}
where, in the last step, we wrote the temperature perturbation as \( \Theta(\vec{k}, {\hat p}, {\hat A})=T(\vec{k}, {\hat p}, {\hat A})\mathcal{R}(\vec{k})\) and used Eq. \eqref{eq:primordials}. In the VDFM model the transfer function depends on ${\hat p}$ and ${\hat A}$ through $\mu$ and ${\hat k}\cdot{\hat A}=\cos\gamma$ respectively, so we expand this function in terms of the Legendre polynomials of $\mu$ 
\begin{equation} \label{eq:transfer_legendre}
    T(k, \mu, \gamma) = \sum_{\ell} (-i)^\ell (2\ell+1) \mathcal{P}_\ell(\mu) T_\ell (k, \gamma).
\end{equation}
A parameterization for the dependence of $T_\ell(k,\gamma)$ with respect to $\g$ has been found to be
\begin{equation}\label{eq:param_transfer}
    T_{\ell}(k,\gamma) =  T_{\ell}(k,\pi/2) + [T_{\ell}(k,0)-T_{\ell}(k,\pi/2)]\cos(\gamma)^2.
\end{equation}
In Fig. \ref{fig:transfer_f} we show this parameterization  together with the transfer functions obtained from \texttt{CLASS} for  $\ell=0$ and different values of $\g$.

Introducing Eqs. \eqref{eq:transfer_legendre} and \eqref{eq:param_transfer} in Eq. \eqref{eq:two-point}, and using the property $    \int d\O Y_{\ell m}({\hat p}) \mathcal{P}_{\tilde \ell} ({\hat p}\cdot {\hat k}) = \frac{4 \pi}{2\ell +1} Y_{\ell m}({\hat k}) \delta_{\ell \tilde \ell}$ \cite{Dodelson:2003ft}, we obtain 
\begin{equation}
\begin{aligned}
 \langle &a_{\ell m} a_{\ell^{'} m^{\prime}}^*\rangle= \int \frac{d k\, k^{2}}{(2\pi)^{3}}\, P_{\mathcal{R}}(k)\, (-i)^{ \ell}  (i)^{\ell^\prime}\, 16\pi^{2}
\int d\O_{k}\; \times \\
\Bigg\{
&\Big[ T_{\ell,0} + (T_{\ell,\pi/2} - T_{\ell,0}) \cos^{2}\gamma \Big] \times  \\ 
& \Big[  T_{\ell',0}^{*} + (T_{\ell',\pi/2}^{*}
- T_{\ell',0}^{*}) \cos^{2}\gamma \Big]
\, Y_{\ell m}({\hat k})\, Y^*_{\ell' m'}({\hat k}) \Bigg\},
\end{aligned}
\end{equation}
where we denoted $T_\ell(k,\pi/2) \equiv T_{\ell,\pi/2}$ and similarly for $T_\ell(k,0)$. After some algebraic manipulations, that involve writing the functions $\cos^{2}\gamma$ and $ \cos^{4}\gamma$ in terms of Legendre polynomials and also using that $\mathcal{P}_\ell ({\hat k} \cdot {\hat A}) = \frac{4\pi}{2\ell+1} \sum_{m=- \ell} ^{ \ell}  Y_{\ell m}({\hat k}) Y^*_{\ell m} ({\hat A})$, we find that the full covariance matrix of temperature anisotropies is given by
\begin{equation}\label{eq:correlator}
\begin{aligned}
&\langle a_{\ell  m}\, a_{\ell ' m'}^{*} \rangle
=
4 \pi \int \frac{dk}{k}
\, \mathcal{P}_{\mathcal{R}}(k)\,
 (-i)^{ \ell}  (i)^{\ell ^\prime}\, 
\Bigg\{
| T_{\ell ,0} |^{2}
\delta_{\ell \ell '} \delta_{mm'}
 \\
&+\frac{1}{35}
( T_{\ell ,\pi/2} - T_{\ell ,0} )
( T_{\ell ',\pi/2}^{ *} - T_{\ell ',0}^{ *} )
I_1 \\
&+\frac{1}{3}\Big[ T_{\ell ,0}
( T^*_{\ell ,\pi/2} - T^*_{\ell ,0} )+
T_{\ell ',0}^{ *} (T_{\ell ',\pi/2}
-T_{\ell ',0}) \Big]\, I_{2}
\Bigg\},
\end{aligned}
\end{equation}
where $I_1$ and $I_2$ are integrals of three spherical harmonic functions that can be written in terms of the Wigner $3j$ symbols as \footnote{The $3j$ Wigner symbols satisfy $\int d \Omega_k Y_{lm}(\hat k)  Y_{\tilde l \tilde m}(\hat k) Y_{l'm'}(\hat k)  = \sqrt{\frac{(2l+1)(2\tilde l+1)(2l'+1)}{4\pi}}\begin{pmatrix}
l &  \tilde l & l'  \\
0 & 0 & 0  \\
\end{pmatrix} \begin{pmatrix}l  & \tilde l& l'  \\m  & \tilde m & m' \\\end{pmatrix}$}
\begin{equation}
\begin{aligned}
I_1 &=8\ (-1)^{m^{\prime}} \sqrt{(2 \ell +1)(2  \ell^\prime+1)} 
\begin{pmatrix}
\ell  & 4 & \ell '  \\
0 & 0 & 0  
\end{pmatrix}
\begin{pmatrix}
\ell  & 4 & \ell '  \\
m & 0 & -m'  
\end{pmatrix}\\
&+20\ (-1)^{m^{\prime}} \sqrt{(2 \ell +1)(2 \ell  ^{\prime}+1)} \begin{pmatrix}
\ell  & 2 & \ell '  \\
0 & 0 & 0  
\end{pmatrix} \begin{pmatrix}
\ell  & 2 & \ell ' \\
m & 0 & -m' 
\end{pmatrix} \\
&+7 \ \delta_{\ell   \ell^\prime} \delta_{m m^\prime},
\end{aligned} \label{eq:I_1}
\end{equation}
\begin{equation}
\begin{aligned}
    I_2 &=2 \ (-1)^{m^{\prime}}\sqrt{(2\ell +1)(2\ell ^\prime +1)} \begin{pmatrix}
\ell  & 2 & \ell '  \\
0 & 0 & 0  
\end{pmatrix}\begin{pmatrix}
\ell  & 2 & \ell '   \\
m & 0 & -m'  \\
\end{pmatrix}\\
&+ \delta_{\ell  \ell ^\prime} \delta_{m m^\prime}. \label{eq:I_2}
\end{aligned}
\end{equation}

The Wigner $3j$ symbols are only non-vanishing if the following rules are satisfied: $m-m'=0$, $|\ell -\ell _0|\leq \ell '\leq \ell +\ell _0$, and $|m|\leq \min (\ell ,\ell ')$. In this case, the first term of $I_1$ is non-zero if $\ell '\in \{ \ell , \ell \pm2, \ell \pm4\}$, while the second term, and also $I_2$, are non-vanishing if $\ell '\in \{ \ell , \ell \pm2 \}$. This means that, the off-diagonal structure is not arbitrary but follows a well-defined angular pattern dictated by the symmetry breaking induced by the background vector field. 

The second and third terms of Eq.~\eqref{eq:correlator} encode the anisotropic contribution of the vector field, as they originate from the $\g$-dependence of the transfer function and vanish in the isotropic limit $T_\ell(k,\g)=T_\ell(k)$.
Figure~\ref{fig:transfer_f} shows the transfer function and the parametrization given by
Eq.~\eqref{eq:param_transfer} for four values of $\g$, considering $\ell =0$ and
$\mA=10^{-24}$ eV. The bottom panel displays the relative difference between
$T_{\pi/2}(k)$ and $T_0(k)$. Over the range
$10^{-4}\lesssim k/(\mathrm{h\,Mpc^{-1}})\lesssim10^{-1}$, this difference is on average of order $10^{-4}$, indicating that the angular dependence of the transfer function is
extremely weak on scales relevant for CMB observations. The multipole range probed by
Planck corresponds approximately to wavenumbers
$k\lesssim0.15\,\mathrm{Mpc}^{-1}$ \cite{Planck:2019nip}, which lies entirely within
this regime. Motivated by this result, in the CMB analyses presented in Sec.~\ref{sec:results} we
neglect the second and third terms of Eq.~\eqref{eq:correlator} and retain only the leading
contribution given by the first term. The validity of this approximation is assessed
quantitatively in App.~\ref{app:delta_Cl}, where we perform a Fisher information analysis of the
anisotropic contributions. For parameter values corresponding to the best-fit
model and to the 95\% confidence bounds obtained under the isotropic approximation, we find that the anisotropic contributions remain well below the sensitivity of an ideal cosmic-variance-limited CMB experiment. This supports
the neglect of these terms in the present analysis.



\section{Methodology and Datasets}\label{sec:methodology}

This section presents the observational datasets and the statistical methodology used to constrain the proposed cosmological scenarios within the vector field dark matter framework introduced above. We begin by describing the main datasets employed in this analysis:

\begin{enumerate}

\item {\bf Cosmic Microwave Background (CMB):}
We use measurements of temperature and polarization anisotropies of the CMB power spectra from the Planck 2018 release, including their cross-spectra and CMB lensing data~\cite{Planck:2018vyg}. In particular, we adopt the high-$\ell$ \texttt{Plik} likelihood for TT (covering $30 \leq \ell \leq 2508$), as well as TE and EE ($30 \leq \ell \leq 1996$), together with the low-$\ell$ TT-only ($2 \leq \ell \leq 29$) likelihood and the low-$\ell$ EE-only ($2 \leq \ell \leq 29$) \texttt{SimAll} likelihood~\cite{Planck:2019nip}. The CMB lensing signal is reconstructed from the temperature four-point correlation function; this likelihood covers the multipole range $8\leq L \leq 400$ ~\cite{Planck:2018lbu}. We refer to this combined dataset as \texttt{Planck}.

\item \textbf{Baryon Acoustic Oscillations} (\textbf{DESI-DR2}):
We include baryon acoustic oscillation (BAO) measurements from the second data release of the Dark Energy Spectroscopic Instrument (DESI). This dataset comprises BAO signals derived from galaxy and quasar samples~\cite{DESI:2025zgx}, as well as from Lyman-$\alpha$ forest tracers~\cite{DESI:2025zpo}. The measurements, summarized in Table~IV of Ref.~\cite{DESI:2025zgx}, cover the redshift range $0.295 \leq z \leq 2.330$ and are divided into nine redshift bins. The BAO observables are expressed in terms of the transverse comoving distance $D_{\mathrm{M}}/r_d$, the Hubble distance $D_{\mathrm{H}}/r_d$, and the volume-averaged distance $D_{\mathrm{V}}/r_d$, all normalized by the sound horizon at the baryon drag epoch, $r_d$. Correlations between these quantities are fully accounted for through the covariance matrix, including the cross-correlation coefficients $r_{\rm V,M/H}$ and $r_{\rm M,H}$. Throughout this work, we denote this dataset as \texttt{BAO (DESI DR2)}.

\item \textbf{Type Ia Supernovae} (\textbf{SNIa}):
We consider the PantheonPlus sample~\cite{Brout:2022vxf}, which includes 1701 light-curve measurements of 1550 distinct supernovae over the redshift range $0.01 \leq z \leq 2.26$ (denoted as \texttt{PP}).

\end{enumerate}
We restrict our analysis to the linear matter power spectrum throughout, since the angular scales probed by \texttt{Planck} data are  within the linear regime; nonlinear corrections (e.g., via the halo model prescription) can be safely neglected for the range of multipoles relevant to our likelihood. For further discussions see Refs.\cite{Verdiani:2026cpc,Gaughan:2026xrv,Hlozek:2017zzf,Lewis:2006fu}. Regarding the inclusion of the lensing likelihood, as shown in App. \ref{app:lensing_vfdm}, the intrinsic anisotropy of the VFDM produces a spurious contribution to the reconstructed lensing potential, corresponding to the $L=2$ multipole, with a much smaller component at $L=4$. Since the Planck lensing likelihood only covers values of $L\geq8$, this spurious contribution falls entirely outside its range; therefore, including it in our analysis requires no additional correction.

We do not include additional late-time geometric probes beyond the combination of $\texttt{BAO (DESI DR2)}$ and PantheonPlus $(\texttt{PP})$, such as alternative Type Ia supernova compilations (e.g., Union~3.0 or DES-Dovekie). At the level of the homogeneous expansion history, VFDM does not introduce meaningful deviations from the $\Lambda$CDM model that would enable these additional geometric datasets to provide qualitatively new constraints. 
Consequently, $\texttt{BAO (DESI DR2)}$ together with $\texttt{PP}$ already provide the relevant late-time distance information for constraining the primary parameter degeneracies in this analysis.

To derive the theoretical predictions of the model and constrain the allowed range of the VFDM particle mass, we modify the \texttt{CLASS} Boltzmann solver~\cite{CLASS_ref} (with details provided in~\cite{Chase:2024wsq}) and perform Markov Chain Monte Carlo (MCMC) analyses using the \texttt{MontePython} sampler~\cite{MP_ref1, MP_ref2}.  
The convergence of the chains is assessed using the Gelman--Rubin criterion~\cite{Gelman_1992}, requiring $R - 1 < 10^{-2}$.  
Post-processing of the Markov chains is carried out with the \texttt{GetDist} package\footnote{\url{https://github.com/cmbant/getdist}}, which is used to extract numerical results, including one-dimensional posterior distributions and two-dimensional marginalized confidence contours.

We perform two complementary analyses. In the first, we assume that the total dark matter component is entirely composed of VFDM. In this case, both the vector field mass and its energy density are treated as free parameters, with flat priors given by $\log_{10}(\mA/\rm{eV}) \in [-26,-19]$ and $\OA \in [0.15,0.3]$,\footnote{From now on, we omit the subscript ``$_{,0}$'' for energy densities evaluated at present time.} respectively.

In the second analysis, we explore a mixed dark matter scenario in which VFDM coexists with standard CDM. This setup is described by two independent density parameters, $\OA$ and $\Ocdm$, each varied with flat priors over the interval $[0,0.5]$. For this case, the prior on the vector field mass is extended to $\log_{10}(\mA/\rm{eV}) \in [-27,-22]$, allowing for a broader exploration of parameter space in the presence of multiple dark matter components.
From the resulting posterior distributions, we compute the fraction of the total dark matter density contributed by VFDM, defined as $f = \tfrac{\OA}{\OA + \Ocdm}$. By construction, this parameter is restricted to the physical interval $0 \leq f \leq 1$, and we explicitly enforce these bounds in the analysis.

In both analyses, we also vary the standard cosmological parameters, namely the physical baryon density $\o_b$, the angular size of the sound horizon at recombination $\theta_*$, the amplitude of primordial scalar perturbations $A_s$, the scalar spectral index $n_s$, and the optical depth to reionization $\tau_{\rm reio}$; along with \texttt{Planck} nuisance parameters. We adopt sufficiently broad flat priors on all these parameters to ensure that the resulting constraints are driven by the data rather than prior assumptions. In particular, we impose the lower bound $\tau_{\rm reio} \geq 0.004$.

We conclude this section by outlining the statistical estimators used to assess the observational viability of the extended cosmological scenarios. We employ two complementary metrics: the minimum effective chi-square, $\chi^2_{\rm min}$, and the Akaike Information Criterion (AIC). The AIC is defined as~\cite{Akaike}
\begin{equation}
{\rm AIC}=-2\ln {\cal L}_{\rm max}+2n_0,
\end{equation}
where ${\cal L}_{\rm max}$ denotes the maximum likelihood and $n_0$ represents the total number of free parameters in the model.

For each model and dataset combination, we evaluate
\begin{equation}
\Delta \chi^2_{\rm min}
=
\chi^2_{\rm min}({\rm our\ model})
-
\chi^2_{\rm min}(\Lambda{\rm CDM}),
\label{eq:dchi2}
\end{equation}
as well as
\begin{equation}
\Delta {\rm AIC}
=
{\rm AIC}({\rm our\ model})
-
{\rm AIC}(\Lambda{\rm CDM}),
\label{eq:daic}
\end{equation}
where $\Lambda$CDM is adopted as the reference cosmological scenario throughout all comparisons. In both estimators, negative values indicate a statistical preference for the extended model under consideration relative to $\Lambda$CDM, while positive values imply that the extended model is statistically disfavored by the data relative to the reference scenario.

\begin{table*}
\centering
\renewcommand{\arraystretch}{1.4}
\setlength{\tabcolsep}{5pt}
\begin{tabular}{l  c  c  c  || c  c}
\hline\hline
& \multicolumn{3}{c||}{VFDM Pure Scenario} & \multicolumn{2}{c}{VFDM + CDM Mixed Scenario} \\
\cline{2-4}\cline{5-6}
Parameter
& CMB
& CMB+BAO
& CMB+BAO+PP
& CMB
& CMB+BAO+PP \\
\hline
$\log_{10}(\mA/\rm{eV})$ [95\% C.L.]
& $>-24.33$
& $>-24.11$
& $>-24.07$
& $>-24.78$
& $>-24.70$ \\
$\Omega_{\rm A}$
& $0.2601^{+0.0068}_{-0.0069}$
& $0.2504^{+0.0023}_{-0.0023}$
& $0.2511^{+0.0024}_{-0.0024}$
& $0.123^{+0.117}_{-0.119}$
& $0.116^{+0.125}_{-0.112}$ \\
$\Omega_{\rm cdm}$
& --
& --
& --
& $0.137^{+0.115}_{-0.115}$
& $0.135^{+0.112}_{-0.123}$ \\
$100\omega_b$
& $2.2379^{+0.0150}_{-0.0152}$
& $2.2518^{+0.0122}_{-0.0120}$
& $2.2508^{+0.0127}_{-0.0119}$
& $2.2386^{+0.0143}_{-0.0145}$
& $2.2503^{+0.0119}_{-0.0121}$ \\
$100\theta_*$
& $1.04186^{+0.00032}_{-0.00031}$
& $1.04203^{+0.00028}_{-0.00027}$
& $1.04203^{+0.00027}_{-0.00027}$
& $1.04185^{+0.00031}_{-0.00030}$
& $1.04201^{+0.00028}_{-0.00028}$ \\
$\ln(10^{10} A_s)$
& $3.0438^{+0.0149}_{-0.0139}$
& $3.0504^{+0.0130}_{-0.0137}$
& $3.0497^{+0.0135}_{-0.0140}$
& $3.0439^{+0.0132}_{-0.0149}$
& $3.0497^{+0.0143}_{-0.0145}$ \\
$n_s$
& $0.9657^{+0.0042}_{-0.0042}$
& $0.9699^{+0.0031}_{-0.0032}$
& $0.9695^{+0.0032}_{-0.0032}$
& $0.9653^{+0.0047}_{-0.0040}$
& $0.9695^{+0.0031}_{-0.0032}$ \\
$\tau_{\rm reio}$
& $0.0538^{+0.0078}_{-0.0072}$
& $0.0588^{+0.0065}_{-0.0066}$
& $0.0583^{+0.0067}_{-0.0066}$
& $0.0539^{+0.0069}_{-0.0078}$
& $0.0584^{+0.0067}_{-0.0075}$ \\
$H_0$
& $67.92^{+0.57}_{-0.57}$
& $68.71^{+0.21}_{-0.20}$
& $68.66^{+0.21}_{-0.21}$
& $67.92^{+0.56}_{-0.56}$
& $68.65^{+0.21}_{-0.21}$ \\
\hline
$\Delta\chi^2_{\rm min}$
& $0.4$
& $-5$
& $-4$
& $-1$
& $-5$ \\

$\Delta$AIC
& $2$ 
& $-3$
& $-2$
& $3$
& $-1$ \\

\hline\hline
\end{tabular}
\caption{Marginalized posterior means for selected cosmological parameters of interest at 68\% CL for the baseline VFDM model, obtained from three dataset combinations for the pure scenario and two
for the mixed scenario\footnote{Here CMB refers to $\texttt{Planck}$ and BAO to $\texttt{BAO (DESI DR2)}$}. For $\log_{10} (m_{\rm A}/\rm{eV})$, we quote instead the 95\% upper limit. In the last two rows, we report the quantities $\Delta \chi^2_{\min} \equiv \chi^2_{\min}(\mathrm{VFDM}) - \chi^2_{\min}(\Lambda\mathrm{CDM})$ and $\Delta \mathrm{AIC} = \mathrm{AIC}_{ \mathrm{min}}(\mathrm{VFDM}) - \mathrm{AIC}_{\mathrm{min}}(\Lambda\mathrm{CDM})$, respectively. }
\label{tab:results}
\end{table*}

\begin{figure}
    \centering
    \includegraphics[width=0.5\textwidth]{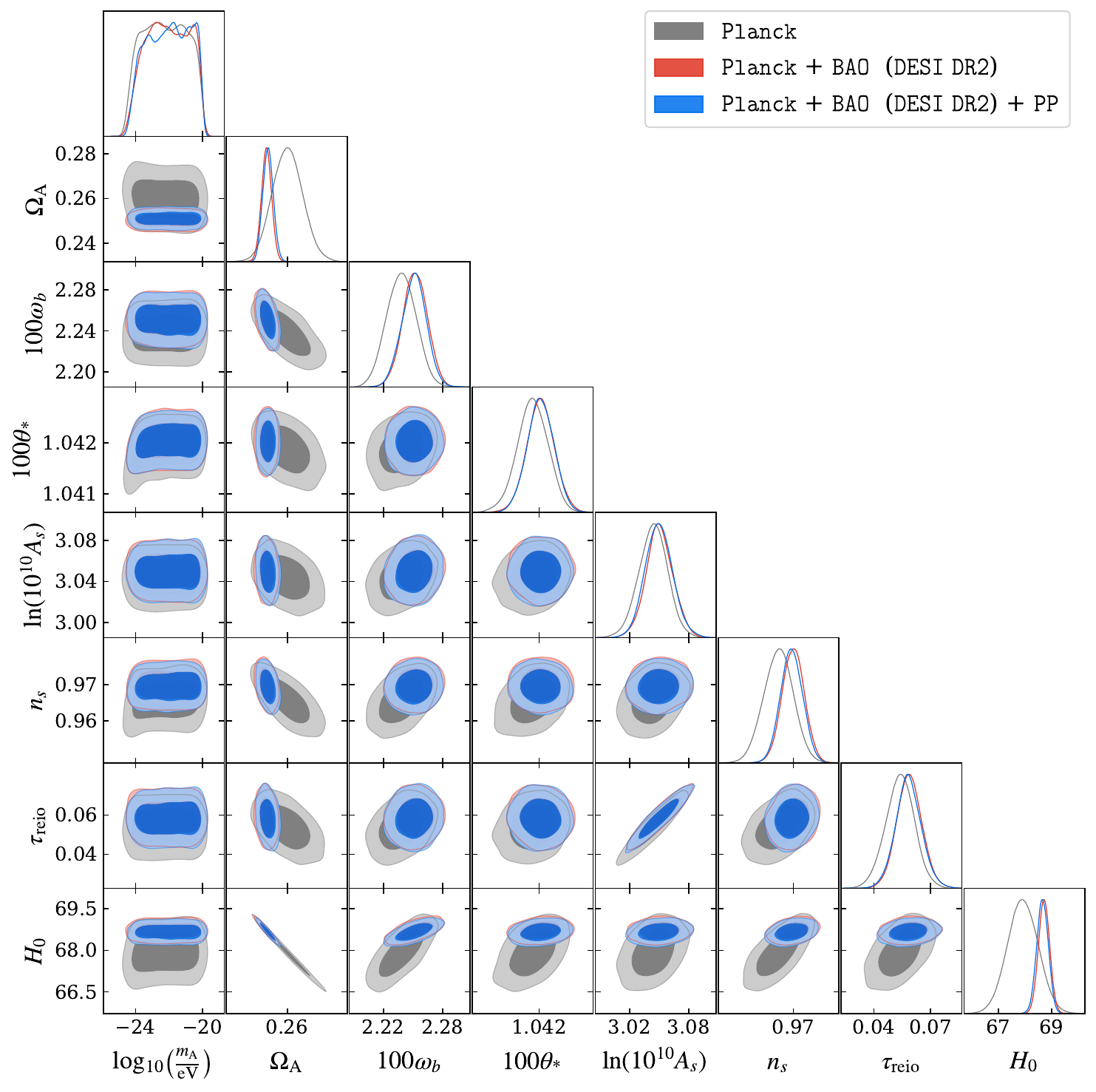}
    \caption{Marginalized posterior distributions for the baseline VFDM model, showing the 68\% and 95\% confidence levels obtained from \texttt{Planck} (gray), \texttt{Planck} + \texttt{BAO (DESI DR2)} (red), and \texttt{Planck} + \texttt{BAO (DESI DR2)} + \texttt{PP} (blue) data combinations. The diagonal panels display the one-dimensional marginalized posteriors, while the off-diagonal panels show the corresponding two-dimensional joint distributions.}
    \label{fig:chains_pure}
\end{figure}

\section{Results}
\label{sec:results}

In what follows, we present and summarize our main results.

\subsection{VFDM Pure Scenario}
The first analysis we make is based on the hypothesis that dark matter consists entirely of an ultralight spin-1 particle described by a vector field, as explained in Sec. \ref{sec:model}. Our aim is to constraint the possible values of this particle's mass by testing the predictions of this theory with three combinations of datasets:  $\texttt{Planck}$, $\texttt{Planck}+\texttt{BAO (DESI DR2)}$, and $\texttt{Planck}+\texttt{BAO (DESI DR2)}+\texttt{PP}$, using MCMC chains according to the methodology described in the previous section. Table~\ref{tab:results} summarizes the main statistical results obtained in our analyses. Figure~\ref{fig:chains_pure} shows the marginalized posterior distributions for the main baseline VFDM model.

The results presented in Table~\ref{tab:results} provide a comprehensive view of the constraints on the VFDM scenario across different dataset combinations. We first highlight the constraints on the vector field mass, which is the key parameter of the model. In all cases, only lower bounds at 95\% C.L. are obtained, namely $\log_{10}(\mA/\rm{eV}) > -24.33$ for \texttt{Planck}, $>-24.11$ for \texttt{Planck}+\texttt{BAO (DESI DR2)}, and $>-24.07$ for \texttt{Planck}+\texttt{BAO (DESI DR2)}+\texttt{PP}. This indicates that current data do not favor a specific mass scale but instead constrain the model to the regime of sufficiently large masses. The inclusion of \texttt{BAO (DESI DR2)} slightly strengthens the bound, reflecting the sensitivity of this parameter to the late-time expansion history and structure formation, while the further addition of \texttt{PP} does not significantly improve the constraint.

The inferred values of the VFDM energy-density parameter, $\OA$, are remarkably stable across the different dataset combinations, converging to $\OA \simeq 0.25$ with progressively smaller uncertainties when $\texttt{BAO (DESI DR2)}$ and $\texttt{PP}$ data are included. This is fully consistent with the total dark matter density inferred within the $\Lambda$CDM framework, indicating that, in the high-mass regime, VFDM effectively reproduces the role of standard CDM.

For the remaining cosmological parameters, we find excellent agreement with the values typically reported in $\Lambda$CDM analyses. The physical baryon density, $100\o_b$, and the angular scale of the sound horizon, $100\theta_*$, are tightly constrained and show negligible shifts relative to their standard values. Similarly, the scalar spectral index, $n_s$, and the amplitude of primordial fluctuations, $\ln(10^{10}A_s)$, remain fully consistent with $\Lambda$CDM expectations, with a mild trend toward slightly higher $n_s$ values when BAO data are included. The optical depth, $\tau_{\rm reio}$, is also stable and in agreement with Planck constraints. The Hubble constant shows a small but consistent shift toward higher values when BAO and supernova data are included, reaching $H_0 \simeq 68.8~\mathrm{km\,s^{-1}\,Mpc^{-1}}$. However, this shift is modest and does not alleviate the Hubble tension relative to local measurements, similarly to what is observed in standard $\Lambda$CDM analyses with the same datasets. 

Finally, the differences in the best-fit chi-square and AIC values indicate that the VFDM model provides a fit to the data that is comparable to that of $\Lambda$CDM, with only modest improvements for the $\texttt{Planck + BAO (DESI DR2)}$ and $\texttt{Planck + BAO (DESI DR2) + PP}$ datasets combinations. In light of these results, we conclude that the model is nearly indistinguishable from the $\Lambda$CDM framework at the background and perturbation levels in the large-scale regime considered here, provided that the VFDM mass satisfies the bound derived here. 


\subsection{VFDM + CDM Mixed Scenario}
We now explore a VFDM + CDM mixed scenario. This setup provides a minimal extension of the VFDM framework, allowing for the coexistence of a standard cold dark matter component with an additional VFDM sector. At the same time, the VFDM contribution can introduce new physical effects through its characteristic mass scale $m_{\rm A}$, particularly at early times. This makes the mixed scenario a phenomenologically viable framework to explore deviations from the pure VFDM and $\L$CDM limits.

\begin{figure} 
    \centering
    \includegraphics[width=\linewidth]{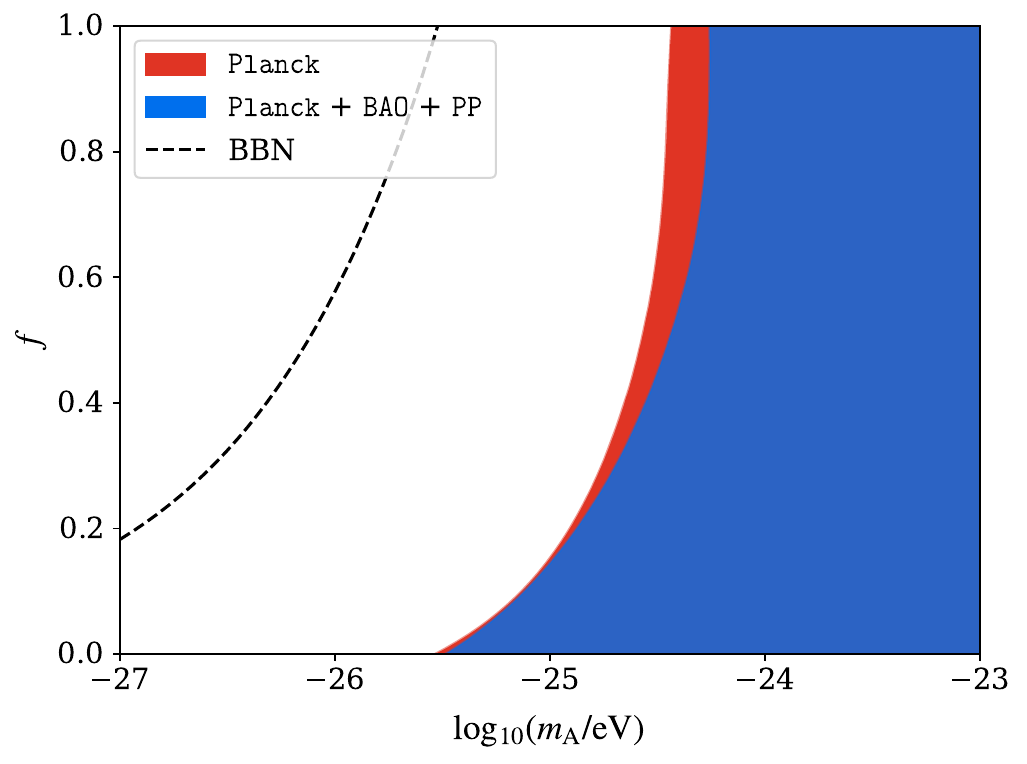}
    \caption{Marginalized $95\%$ contour of the fraction $f$ of VFDM vs $\log_{10}(\mA/\rm{eV})$ obtained from \texttt{Planck} (red), and \texttt{Planck} + \texttt{BAO (DESI DR2)} + \texttt{PP} (blue) data combinations. The BBN constraint, given by Eq. \eqref{eq:BBN_constr}, is shown in the dashed line.}
    \label{fig:m_vs_f_95}
\end{figure}

To perform this analysis, we run MCMC chains similar to the previous ones, but now simultaneously sampling $\Omega_{\rm cdm}$ and $\Omega_{\rm A}$. The right panel of Table~\ref{tab:results} summarizes the main statistical results. Fig.~\ref{fig:m_vs_f_95}, shows the $95\%$ C.L. contour of $f$ versus $\log_{10}(\mA/\rm{eV})$, where we also include the BBN constraint given by Eq. \eqref{eq:BBN_constr}. 

 Figure~\ref{fig:m_vs_f_95} shows a clear correlation between the allowed values of the VFDM fraction $f$ and the vector-field mass $m_{\rm A}$: lower masses are compatible with the data only if the VFDM fraction is sufficiently small, while larger fractions require progressively higher masses. Since  the total dark matter abundance remains fixed, i.e. $\Omega_{\rm A} + \Omega_{\rm cdm} \equiv \Omega_{\rm dm}$, as $f$ decreases, CDM becomes the dominant component driving structure formation and the expansion history, reducing the impact of the vector field on cosmological observables. In other words, the departures from $\L$CDM induced by VFDM become smaller either when the vector field is heavier or when its fractional contribution to the dark matter abundance is reduced. 

This is quantitatively reflected in Table~\ref{tab:results}. In the pure VFDM scenario ($f=1$), the model is required to fully reproduce the role of dark matter, leading to relatively strong lower bounds on the mass, $\log_{10}(\mA/\rm{eV}) \gtrsim -24.07$ (from CMB+BAO+PP). In contrast, in the mixed scenario, where after marginalizing the mass the data give $\Omega_{\rm A} \sim 0.12$  with large uncertainties (consistent with any fraction), the constraints relax to $\log_{10}(\mA/\rm{eV}) \gtrsim -24.7$. This shift highlights the dependence of the bounds on the assumed fraction $f$. 

The BBN constraint, shown as a dashed line in Fig.~\ref{fig:m_vs_f_95}, excludes the region of low masses and moderate-to-large VFDM fractions. It therefore provides an additional lower envelope in the $(f, \log_{10} (m_{\rm A}/\rm{eV}))$ plane, complementing the cosmological bounds derived from the MCMC analysis.

In comparison with the corresponding bounds obtained for the scalar field (see Ref. \cite{Hlozek:2014lca}), the constraint on $f$ in the vector-field case becomes significantly more stringent at low masses, exhibiting a much steeper mass dependence than in the scalar-field case. The origin of this difference lies in the distinct early-time evolution of the two scenarios. Before the onset of oscillations, the vector-field energy density redshifts as radiation, whereas the scalar-field energy density remains approximately constant and becomes negligible at sufficiently early times. Consequently, lowering the vector-field mass enhances its impact on the pre-recombination expansion history, leading to substantially stronger cosmological constraints.

An additional radiation component increases the expansion rate prior to recombination and reduces the sound horizon, resulting in a shift of the $H_0$ value inferred from the CMB toward higher values. Specifically, as shown in Sec. \ref{sec:model}, the amount of radiation increases with lower masses according to the relation $R_A \equiv \frac{\rho_A}{\rho_r}\propto f \, {m_{\rm A}}^{-1/2}$, so in the mixed scenario, in which lower masses are allowed, one would naively expect the well-known Hubble tension to be alleviated to some extent. 
However, the CMB constrains not only the angular size of the sound horizon, $\theta_*$, but also additional characteristic scales, including the damping scale and the scale associated with matter-radiation equality. As the vector field mass is decreased, the enhanced radiation density modifies all of these observables simultaneously. We find that the masses required to produce a significant reduction of the sound horizon are already disfavored by the impact of VFDM on the remaining CMB observables. Consequently, the model does not generate a meaningful shift in the inferred value of $H_0$, and the preferred cosmological parameters remain close to their $\Lambda$CDM values. This behavior is consistent with recent studies showing that successful early-time solutions to the Hubble tension require a specific modification of the pre-recombination expansion history. In particular, such solutions must simultaneously preserve the geometric relations among the sound horizon, damping scale, and equality scale \cite{Uzan:2023dsk,Pitrou:2023swx,Pedrotti:2026dwj}. The VFDM background evolution considered here does not realize this type of pre-recombination modification within the region of parameter space favored by current observations.

\section{Off-diagonal CMB covariance matrix with VDFM}\label{sec:offdiag}

The goal of this section is to quantify the relevance of non-diagonal elements of the CMB covariance matrix in constraining the model.  For this we adopt the best-fit  parameters derived in previous sections and work in the Bipolar Spherical Harmonic (BipoSH) formalism \cite{Hajian:2003qq,Hajian:2005jh}.

The BipoSH formalism captures statistical isotropy violations via coefficients that are an equivalent representation of the spherical harmonic correlation matrix, and are given by 
\begin{equation}
    A^{LM}_{\ell\ell'} = \sum_{m m'} (-1)^{m'} \langle a_{\ell m}a_{\ell' m'}^*\rangle \mathcal{C}^{LM}_{\ell m\ell'-m'}
    \label{A_def}
\end{equation}
where $\mathcal{C}^{LM}_{\ell m\ell'-m'}$ are Clebsch-Gordan coefficients\footnote{We use $\mathcal{C}^{LM}_{\ell m\ell'm'}=(-1)^{\ell-\ell'+M} \sqrt{2L+1}\begin{pmatrix}
\ell & \ell' & L\\
m & m' & -M
\end{pmatrix}$.}.
The usual  CMB angular power spectrum corresponds to fixing $L=0$, obtaining $A^{00}_{l l^{\prime}}  = (-1)^l \sqrt{2l+1} C_{l} \delta_{l l^{\prime}}$.
The  coefficients with $L>0$ quantify violations of statistical isotropy. In the standard $\Lambda$CDM scenario they vanish due to the isotropy of the background. In the VFDM model, we obtain non-vanishing coefficients for $L=2,4$. 
We thus focus on these quantities.

By inserting the expression for the covariance matrix given in Eq. \eqref{eq:correlator} into Eq. \eqref{A_def}, and using the  orthogonality relation of the Wigner $3j$ symbols, 
\begin{equation}\label{OrthogWS}
    \sum_{m_1, m_2}(2 j+1)\left(\begin{array}{ccc}j_1 & j_2 & j \\ m_1 & m_2 & m\end{array}\right)\left(\begin{array}{ccc}j_1 & j_2 & j^{\prime} \\ m_1 & m_2 & m^{\prime}\end{array}\right)=\delta_{j  j^{\prime}} \delta_{m m^{\prime}},
\end{equation}
we obtain  
\begin{align}
  A^{2,0}_{\ell\ell'} \;=\; &\frac{8\pi}{3\sqrt{5}}\sqrt{(2\ell+1)(2\ell'+1)}\,(-i)^{\ell'}i^{\ell}\begin{pmatrix} \ell & 2 & \ell' \\ 0 & 0 & 0 \end{pmatrix} \label{A_expression}
  \\&\times\left[G_\ell+G_{\ell'} + \frac{6}{7}\, J_{\ell\ell'} \right],\nonumber
\end{align}
for $\ell'\in\{\ell,\ell\pm2\}$, and
\begin{align}
  A^{4,0}_{\ell\ell'} \;=\; &\frac{32\pi}{105}\sqrt{(2\ell+1)(2\ell'+1)}\,(-i)^{\ell'}i^{\ell}\begin{pmatrix} \ell & 4 & \ell' \\ 0 & 0 & 0 \end{pmatrix} \label{A_expression}\, J_{\ell\ell'},
\end{align}
for $\ell'\in\{\ell,\ell\pm2, ,\ell\pm4\}$, with
\begin{equation}
  G_\ell
  = \int_0^\infty \frac{dk}{k}\, \mathcal{P}_\mathcal{R}(k)\,
    \, T_{\ell,0}(k)\big(T_{\ell,\pi/2}(k)-T_{\ell,0}(k)\big)\, ,
\end{equation}

\begin{align}
  J_{\ell\ell'}
  = \int_0^\infty \frac{dk}{k}\, \mathcal{P}_\mathcal{R}&(k)\,\big(T_{\ell,\pi/2}(k)-T_{\ell,0}(k)\big)\\
    &\times\big(T_{\ell',\pi/2}(k)-T_{\ell',0}(k)\big)\,. \nonumber
\end{align}

\begin{figure}
\centering
    \includegraphics[width=0.95\linewidth]{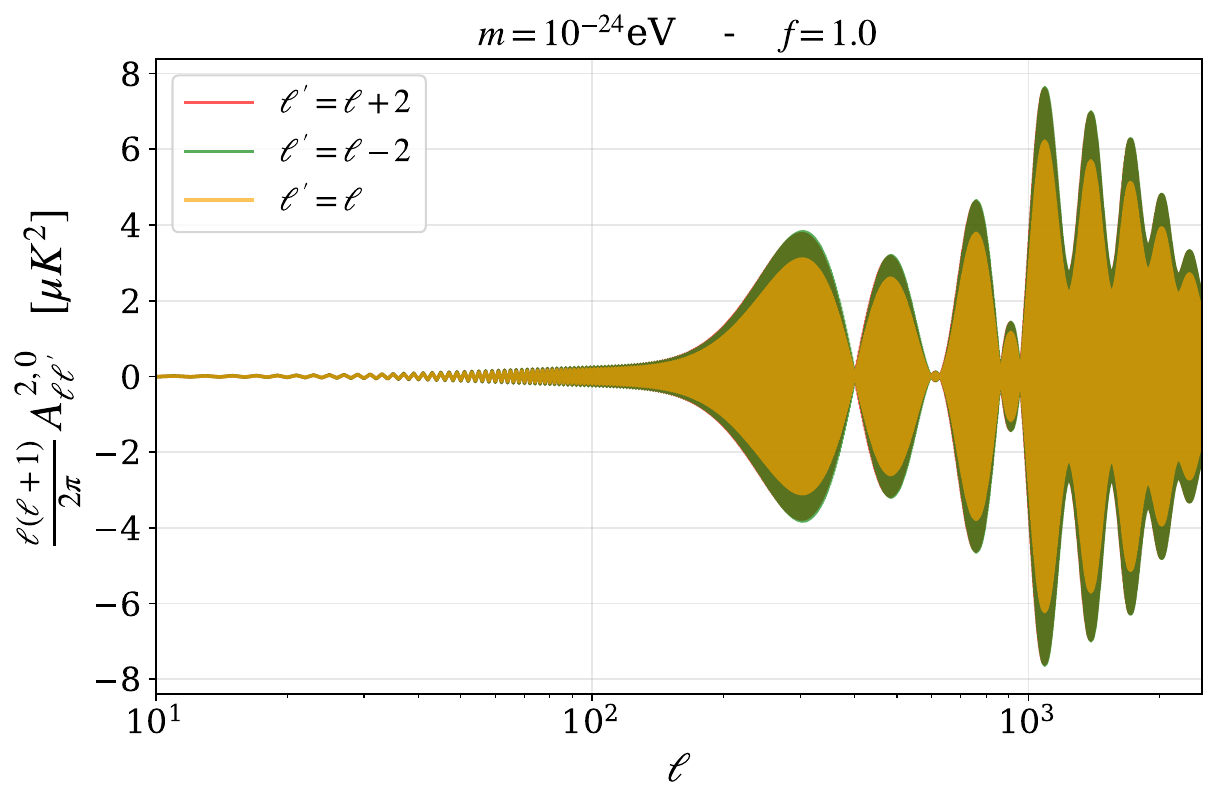}\vfill
    \includegraphics[width=0.95\linewidth]{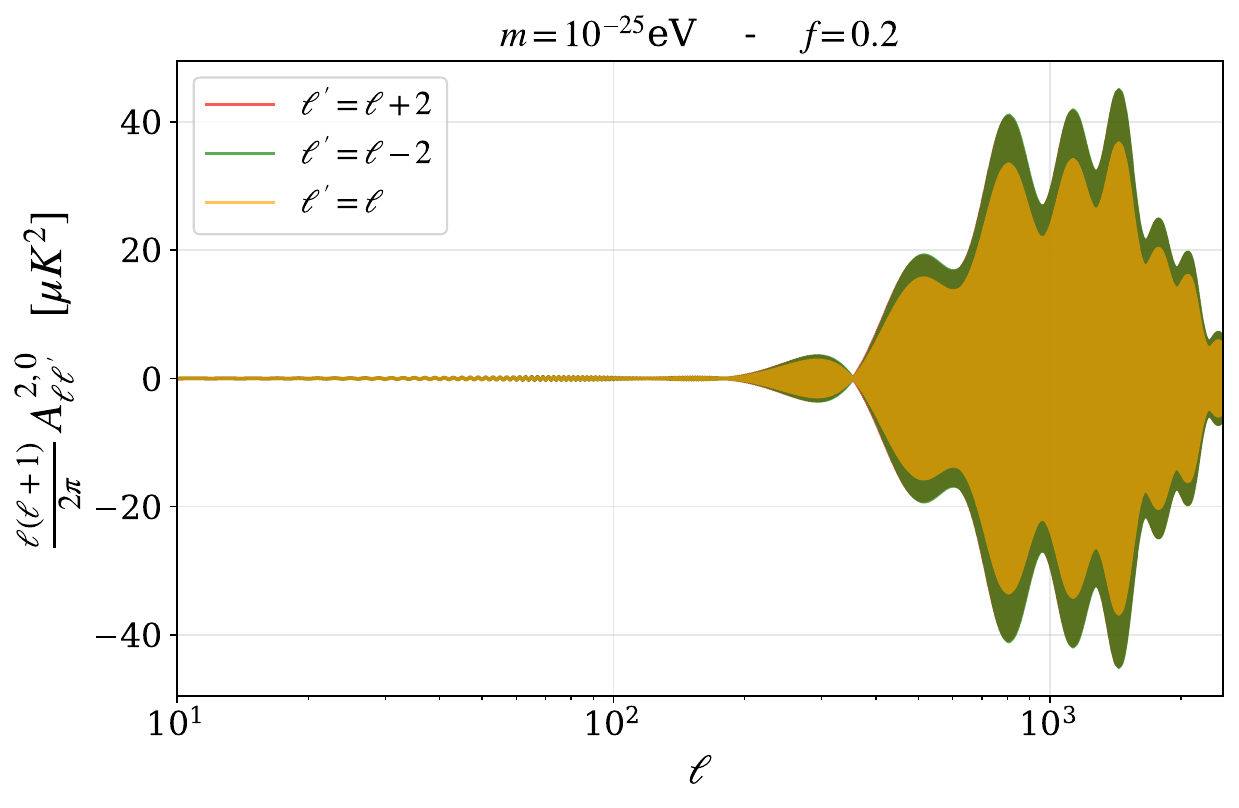} 
    \caption{BipoSH coefficients $A^{20}_{\ell,\ell}$ and $A^{20}_{\ell,\ell\pm2}$ for the VFDM+CDM model,  using our best-fit for the cosmological parameters (Table \ref{tab:results}), and different masses and fractions of the vector field near the contour of Fig. \ref{fig:m_vs_f_95}. Notice the  curve corresponding to $\ell'=\ell+2$ (in red) is barely visible because it is almost completely overlapped by the one for $\ell'=\ell-2$ (in green).  }
    \label{fig:biposh_power_spectrum_20}
\end{figure}

In Fig. \ref{fig:biposh_power_spectrum_20} we show the predictions for $A^{20}_{\ell,\ell'}$ and $A^{20}_{\ell,\ell'\pm2}$ with  VFDM, calculated using our best-fit for the cosmological parameters (Table \ref{tab:results}),  different masses of the vector field and fractions $f$, which lie near  the contour shown in Fig. \ref{fig:m_vs_f_95}. These predictions can be compared with the BipoSH coefficients extracted from \texttt{Planck} 2013 maps (see Fig. 38 in Ref. \cite{Planck:2013lks}). In particular, we can see that the predicted coefficients in the VFDM model (with $f=1$) lie within   the error bars of the coefficients  extracted from the map, which are themselves consistent with zero at the  $3\sigma$ level. This indicates that including estimators of these coefficients is unlikely to yield tighter constraints on the parameter space  than those already obtained from the diagonal part of the covariance matrix. However, for a VFDM + CDM model with $f=0.2$ the magnitude becomes comparable with Planck sensitivity, and even lies above it for certain multipoles. This motivates a dedicated analysis of mixed VFDM + CDM scenarios that incorporates the off-diagonal components of the temperature covariance matrix, which could either tighten the current constraints on the VFDM fraction and mass or enable the detection of the characteristic anisotropic signatures induced by the vector field. 

\begin{figure}
    \centering
    \includegraphics[width=0.95\linewidth]{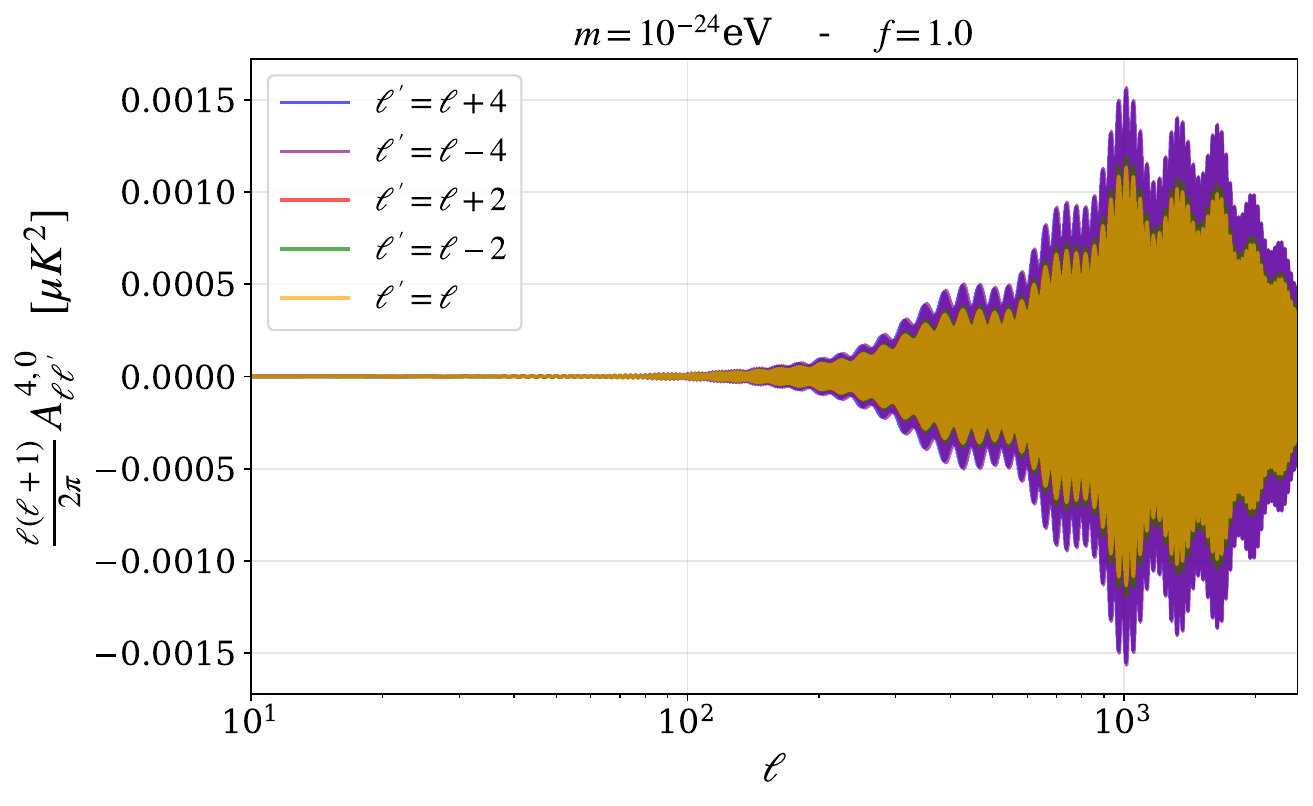}
    \includegraphics[width=0.95\linewidth]{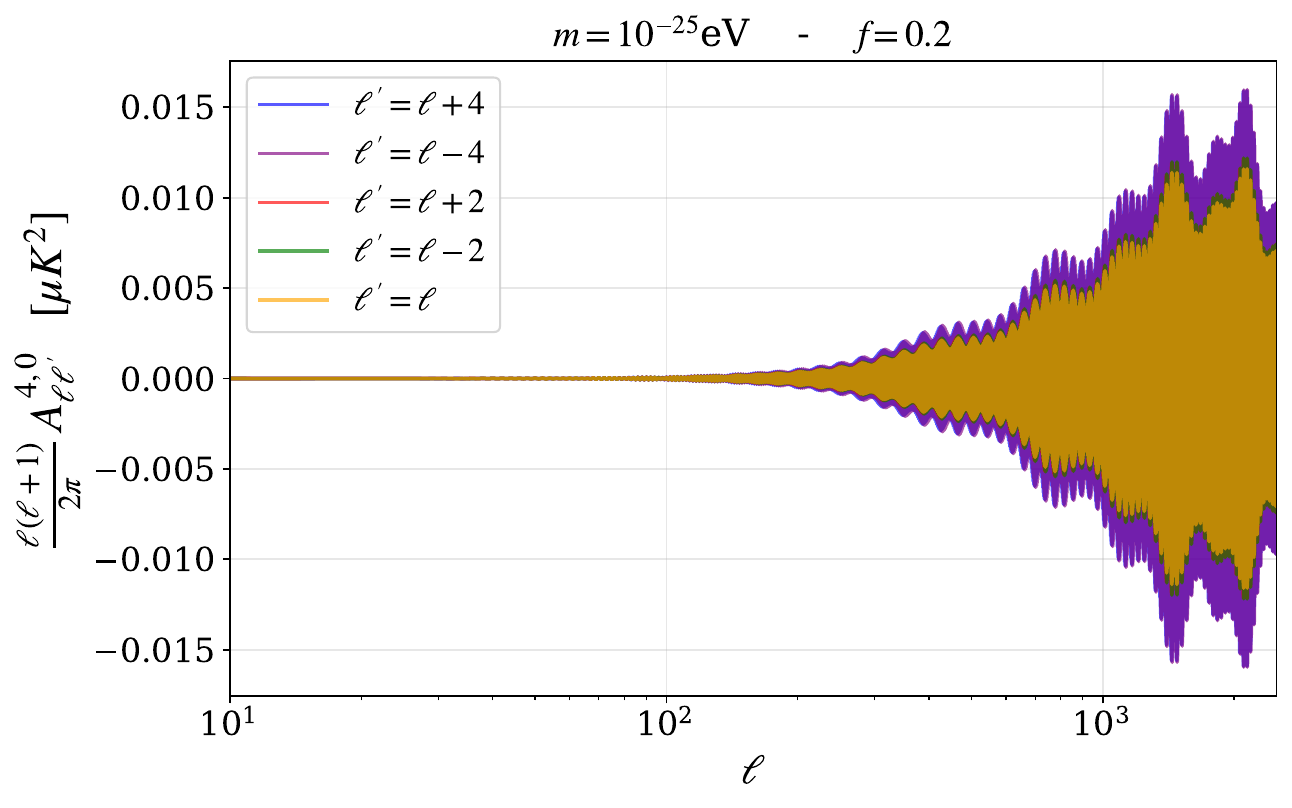}
    \caption{BipoSH coefficients $A^{40}_{\ell,\ell}$ and $A^{40}_{\ell,\ell\pm2}$ for VFDM+CDM scenario  with the same parameters as in Fig.~\ref{fig:biposh_power_spectrum_20}.}
    \label{fig:biposh_power_spectrum_40}
\end{figure}

For the sake of completeness, in Fig. \ref{fig:biposh_power_spectrum_40} we show the predictions of the BipoSH coefficients for $L=4$, which are much smaller than the ones for  $L=2$.


\section{Conclusions}\label{sec:conclusions}
In this work we have presented the first systematic Bayesian analysis of ultralight VFDM  with anisotropic background using current cosmological observations. We considered a spin-1 dark matter candidate described by a massive Proca field and constrained its parameter space through Markov Chain Monte Carlo analyses combining Planck 2018 CMB measurements with recent BAO data from DESI DR2 and Type Ia supernovae from the PantheonPlus compilation.

On the theoretical side, we derived the full covariance matrix of CMB temperature anisotropies in the VFDM framework. Due to the anisotropic Bianchi-I background sourced by the vector field, the model predicts both diagonal and off-diagonal correlations. However, we showed that these anisotropic contributions are strongly suppressed on the scales probed by current Planck CMB observations for the region of parameter space compatible with the data. As a consequence, the dominant observational signatures originate from the modifications of the isotropic part of the transfer functions and the background expansion history. We further proved that the intrinsic anisotropy of the background vector field generates spurious contributions that could bias the inferred lensing signals. Nonetheless, these contributions correspond to very low multipoles, $L=2$ and $L=4$, which fall outside the range currently probed by the datasets.  

For the pure VFDM scenario, in which the vector field constitutes the entire dark matter abundance, we obtained a robust lower bound $\log_{10}(m_{\rm A}/{\rm eV}) > -24.07$ at $95\%$ confidence level from our most constraining dataset combination. This result indicates that current cosmological observations push the model toward the regime where the vector field begins oscillating early enough and behaves effectively as standard cold dark matter throughout most of cosmic history.  Correspondingly, the standard cosmological parameters remain fully consistent with their $\Lambda$CDM values, and the overall statistical performance of the model is comparable to that of the concordance cosmology.

In the mixed VFDM+CDM scenario, we find a clear correlation between the vector field fraction $f$ and its mass, with smaller fractions allowing significantly lighter vector masses. Compared to scalar ultralight dark matter models, the resulting constraints become substantially more stringent at low masses. This difference originates from the distinct early-time evolution of the two scenarios: before oscillations, the vector-field energy density scales as radiation and therefore contributes non-negligibly to the total energy budget of the early Universe, whereas the scalar-field energy density remains approximately constant and becomes negligible at sufficiently early times.

In this work we restricted our analysis to the linear matter power spectrum, consistent with the range of scales probed by the \texttt{Planck} dataset. Extending this analysis to surveys such as \texttt{ACT} \cite{AtacamaCosmologyTelescope:2025nti}, which reconstruct CMB lensing from smaller, mildly nonlinear scales, would require modeling these nonlinear corrections, for instance within the effective field theory of large-scale structure \cite{Verdiani:2025jcf} or by using prescriptions like the halo model \cite{Gaughan:2026xrv}. Beyond this, the off-diagonal structure of the CMB covariance predicted by VFDM provides a distinctive signature of the model. As discussed in the previous section, the associated BipoSH coefficients that characterize this structure are challenging to detect with Planck CMB data in the case of pure VFDM model. However, we have shown that mixed VFDM + CDM scenarios can produce anisotropic signatures at the level of, or even above, the Planck sensitivity over a range of multipoles, motivating a dedicated data analysis. Future work in this direction could also investigate whether improved CMB data, including a more accurate modeling of secondary anisotropies, together with dedicated estimators designed to probe this particular non-diagonal covariance structure, could enable a more sensitive search for this signature.  Moreover, complementary probes of large-scale structure, such as Ly$\alpha$ forest measurements and future 21 cm observations, could provide additional avenues to constrain the VFDM model or detect its anisotropic signatures. 

Overall, our results provide the first comprehensive cosmological constraints on spin-1 ultralight dark matter. The resulting constraints define the currently allowed parameter space for mixed spin-1 ultralight dark matter models and significantly improve upon simple background-level estimates based on Big Bang nucleosynthesis. Although the current data slightly favor the VFDM model over the $\Lambda$CDM model (with modest improvements in the best-fit chi-square and AIC values for the $\texttt{Planck + BAO (DESI DR2)}$ and $\texttt{Planck + BAO (DESI DR2) + PP}$ datasets combinations), they do significantly constrain its space of viable parameters and highlight the important role that observables from the early Universe play in testing this class of models.
 

\begin{acknowledgments}
\noindent G.A.A. gratefully acknowledges the Astronomy Department at UFRGS, where most of this work was developed under CAPES/AUGM project No. 88881.004428/2024-01. G.A.A., T.F.C. and D.L.N. were supported by  CONICET and UBA.  R.C.N. acknowledges financial support from the Conselho Nacional de Desenvolvimento Científico e Tecnológico (CNPq, National Council for Scientific and Technological Development) through Grant No. 304306/2022-3, and partial financial support from the Fundação de Amparo à Pesquisa do Estado do Rio Grande do Sul (FAPERGS, Research Support Foundation of the State of Rio Grande do Sul) through Grant No. 23/2551-0000848-3 and Grant No. 25/2551-0002612-1.  
\end{acknowledgments}

\appendix
\section{Detectability of the anisotropic terms of Eq. \eqref{eq:correlator} on the angular power spectrum} \label{app:delta_Cl}
In this appendix we assess the observability of the anisotropic ($\gamma$-dependent) contributions appearing in the diagonal part of Eq.~\eqref{eq:correlator}, for an idealized (cosmic-variance-limited) measurement of the angular power spectrum. To this end, we introduce an auxiliary amplitude parameter $g$ through
\begin{equation}
\langle a_{\ell m} a_{\ell' m'}^* \rangle \Big |_{\rm diag}
=
C_\ell\,
\delta_{\ell\ell'}
\delta_{mm'}
+
g\,\delta C_{\ell m,\ell m},
\end{equation}
where $\delta C_{\ell m,\ell m}$ contains all the diagonal ($\ell=\ell', m=m'$), anisotropic terms proportional to $(T_{\ell,0}-T_{\ell,\pi/2})$. The physical prediction of the model corresponds to $g=1$, while $g=0$ recovers the isotropic limit.

The likelihood analysis presented in this work is based on the averaged angular power spectrum. Therefore, the estimator used in the MCMC chains is
\begin{equation}
 \hat C_\ell
=
\frac{1}{2\ell+1}
\sum_{m=-\ell}^{\ell}
|a_{\ell m}|^2 .
\end{equation}
The relevant anisotropic correction is then
\begin{equation}\label{eq:app_estimator_Cl}
\hat{\delta C_\ell}
=
\frac{1}{2\ell+1}
\sum_{m=-\ell}^{\ell}
\delta C_{\ell m,\ell m}\,.
\end{equation}
For an ideal cosmic-variance-limited experiment, the variance of the power-spectrum estimator is
\begin{equation}
\mathrm{Var}(\hat C_\ell)
=
\frac{2}{2\ell+1}
\hat C_\ell^2 \, .
\end{equation}
The Fisher information associated with the auxiliary parameter $g$ is given by \cite{Dodelson:2003ft}
\begin{equation}
F_{gg}^{(\hat C_\ell)}
=
\frac12
\sum_\ell
(2\ell+1)
\frac{\left(\hat{\delta C_\ell}\right)^2}
     {\hat C_\ell^2}.
\end{equation}

According to the Cramér--Rao bound, the minimum uncertainty with which $g$ can be measured is \cite{Tegmark:1996bz}
\begin{equation}
\sigma_g
=
\frac{1}{\sqrt{F_{gg}}}.
\end{equation}
Since $F_{gg}^{(C_\ell)}$ is evaluated assuming vanishing instrumental
noise, $\sigma_g$ should be interpreted as the most
optimistic bound that can be achieved observationally.
Given that the theory predicts $g=1$, a necessary condition for its detection is
$\sigma_g < 1$. Otherwise, the minimum statistical uncertainty would exceed the signal amplitude itself. The resulting values of $F_{gg}^{(C_\ell)}$ and $\sigma_g$ are shown in Table~\ref{tab:fisher_detectability}, using the best-fit parameters obtained from the CMB analysis of Sec.~\ref{sec:results}. The pairs $(f,\log_{10} (m_{\rm A}/\rm{eV}))$ correspond to representative points along the $95\%$ confidence contour (see Fig. \ref{fig:m_vs_f_95}).

\begin{table}[h]
\centering
\begin{tabular}{l|c|c}
\hline\hline
 $(f,\log_{10} (m_{\rm A}/\rm{eV}))$ &
$F_{gg} $ & $\sigma_{g}$\\
\hline

$(0.01,-25.5)$ &
$2\times10^{-2}$ &
$8$ \\

$(0.10,-25.2)$ &
$1\times10^{-1}$ &
$3$ \\

$(0.50,-24.6)$ &
$5\times10^{-1}$ &
$1.4$ \\

$(1.00,-24.4)$ &
$3\times10^{-4}$ &
$1.1$  \\

\hline\hline
\end{tabular}
\caption{Fisher information and Cramér--Rao uncertainties for some pairs $(f,\log_{10} (m_{\rm A}/\rm{eV}))$ along the $95\%$ confidence contour in parameter space.}
\label{tab:fisher_detectability}
\end{table}
In all cases we find $\sigma_g > 1$, implying that the minimum statistical uncertainty is larger than the amplitude predicted by the model ($g=1$). This result provides a quantitative justification for neglecting the diagonal, anisotropic contributions in Eq.~\eqref{eq:correlator} and retaining only the leading isotropic term in the CMB analysis presented in Sec.~\ref{sec:results}.

However, an analysis based on the full covariance matrix $\langle a_{\ell m} a_{\ell'm'} \rangle$ (rather than on an average, as in Eq. \eqref{eq:app_estimator_Cl}) could provide a more sensitive probe of the anisotropic signatures predicted by the model, including the off-diagonal contributions. We leave such an analysis for future work.

\section{Lensing bias from the statistical anisotropy of VFDM}\label{app:lensing_vfdm}

The quadratic estimator for the lensing potential can be written as \cite{Planck:2013mth, Okamoto:2003zw}
\begin{equation}
\hat\phi_{LM} = N_L
\sum_{\substack{\ell_1 m_1,\\ \ell_2 m_2}}
(-1)^M
\begin{pmatrix}
\ell_1 & \ell_2 & L\\
m_1 & m_2 & -M
\end{pmatrix}
g_{\ell_1\ell_2L}\,
\frac{a_{\ell_1 m_1}}{C_{\ell_1}^{\rm tot}}
\frac{a_{\ell_2 m_2}}{C_{\ell_2}^{\rm tot}} , \label{eq:app_QE}
\end{equation}
where $g_{\ell_1\ell_2 L}$ is the standard lensing response kernel, $N_L$ is the normalization function and $C_{\ell}^{\rm tot}$ is the lensed CMB power spectrum including instrumental noise.

In VFDM, the background vector field selects a preferred direction $\hat A$, which induces an additional off-diagonal contribution to the covariance,
\begin{equation}
\langle a_{\ell m} a^*_{\ell' m'}\rangle = C_\ell\,\delta_{\ell\ell'}\delta_{mm'} + \Delta C^{\rm VFDM}_{\ell m,\, \ell' m'} + \left(\text{lensing}\right),
\label{eq:app_cov}
\end{equation}
with $\Delta C^{\rm VFDM}_{\ell m,\,\ell' m'}$ given explicitly by Eqs.~(\ref{eq:correlator})–(\ref{eq:I_2}) of the main text. Therefore, taking the expectation value of $\hat\phi_{LM}$ (Eq. \eqref{eq:app_QE}) and using Eq. \eqref{eq:app_cov} and the property $\langle a_{\ell m} a_{\ell' m'}^*\rangle = (-1)^{m'} \langle a_{\ell m} a_{\ell' -m'} \rangle$, we find
\begin{equation} 
\begin{aligned}
\langle\hat\phi_{LM}\rangle_{\rm bias} = N_L \sum_{\substack{\ell_1 m_1,\\ \ell_2 m_2}}& (-1)^M
\begin{pmatrix}\ell_1 & \ell_2 & L\\ m_1 & m_2 & -M\end{pmatrix}
g_{\ell_1\ell_2 L}\times \\
&\frac{(-1)^{m_2}\Delta C^{\rm VFDM}_{\ell_1 m_1,\, \ell_2 -m_2}}{C_{\ell_1}^{\rm tot} C_{\ell_2}^{\rm tot}} .
\label{eq:app_bias}
\end{aligned}
\end{equation}
This is formally identical to the mean-field calculation performed for the mask or for anisotropic noise (see for instance \cite{Planck:2013mth, Planck:2018lbu}), except that $\Delta C^{\rm VFDM}$ is known analytically rather than obtained from Monte Carlo simulations. In this sense, the vector field act as a spurious mean-field bias to reconstructed lensing.  

We note that $(-1)^{m_2}\Delta C^{\rm VFDM}_{\ell_1 m_1, \ell_2 -m_2}$ contains terms with two different Wigner $3j$ symbols, $ \begin{pmatrix}\ell_1 & 2 & \ell_2\\ m_1 & 0 & m_2 \end{pmatrix}$ and  $\begin{pmatrix}\ell_1 & 4 & \ell_2\\ m_1 & 0 & m_2 \end{pmatrix}$. The first is linear in $\varepsilon_\ell \equiv (T_{\ell,\pi/2}-T_{\ell,0})/T_{\ell,0} \sim 10^{-4}$ over CMB scales, and the second one is quadratic in $\varepsilon_\ell$, being suppressed by $\varepsilon_\ell ^2\sim 10^{-8} $ at those scales. Using the  orthogonality relation of the Wigner $3j$ given in Eq.\eqref{OrthogWS} we find, after using the symmetry properties of the Wigner $3j$ symbols under column permutations, that only the $L=2$ and $L=4$ multipoles contribute to Eq. \eqref{eq:app_bias}, both with $M=0$.
 
The multipoles at which our model predicts a spurious lensing-like contribution ($L=\{2,4\}$) lie entirely outside the range where the Planck 2018 lensing likelihood is constructed ($L\geq8$) \cite{Planck:2018lbu}, so including this dataset in our analysis does not require any additional correction for this effect.

\bibliographystyle{apsrev4-1} 
\bibliography{bibliography} 

\end{document}